\documentclass[a4paper,11pt]{article}
\usepackage{pos}
\usepackage{float}

\title{Evaluation and comparison of trigger strategies for the SWGO}

\author*[a]{Johannes Bennemann}
\onbehalf{on behalf of the SWGO collaboration\footnote[2]{
  Full list of authors available at:
  {\tiny\url{https://www.swgo.org/SWGOWiki/lib/exe/fetch.php?media=wiki:the_swgo_collaboration_ICRC25.pdf}}
}}

\affiliation[a]{Max-Planck-Institut für Kernphysik (MPIK),\\
Saupfercheckweg 1, Heidelberg, Germany}

\emailAdd{johannes.bennemann@mpi-hd.mpg.de}

\abstract{
  The Southern Wide-field Gamma-ray Observatory (SWGO) is a future TeV gamma-ray observatory to be built in Chile. The high-duty cycle experiment will consist of about 4000 water-Cherenkov tanks equipped with two photomultiplier tubes each. The available computational infrastructure will not be sufficient to handle all the raw data SWGO will produce. Therefore, to limit the computational requirements, it is necessary to develop a sophisticated trigger strategy. This trigger strategy reduces the amount of data while also ensuring its scientific quality.
  In this contribution, we present our developed framework to test and validate different trigger strategies. We came up with a metric for the trigger quality, allowing us to compare trigger strategies with each other. We present the developed tools and demonstrate their use.
}

\ConferenceLogo{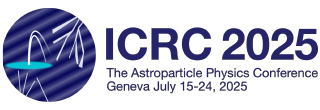}

\FullConference{39th International Cosmic Ray Conference (ICRC2025)\\
 15–24 July 2025\\
Geneva, Switzerland\\}

\begin{document}
\maketitle

\section{Introduction}

\begin{figure}[htb]
  \centering
  \includegraphics[width=0.8\textwidth, trim={0 0 0 3.6cm}, clip]{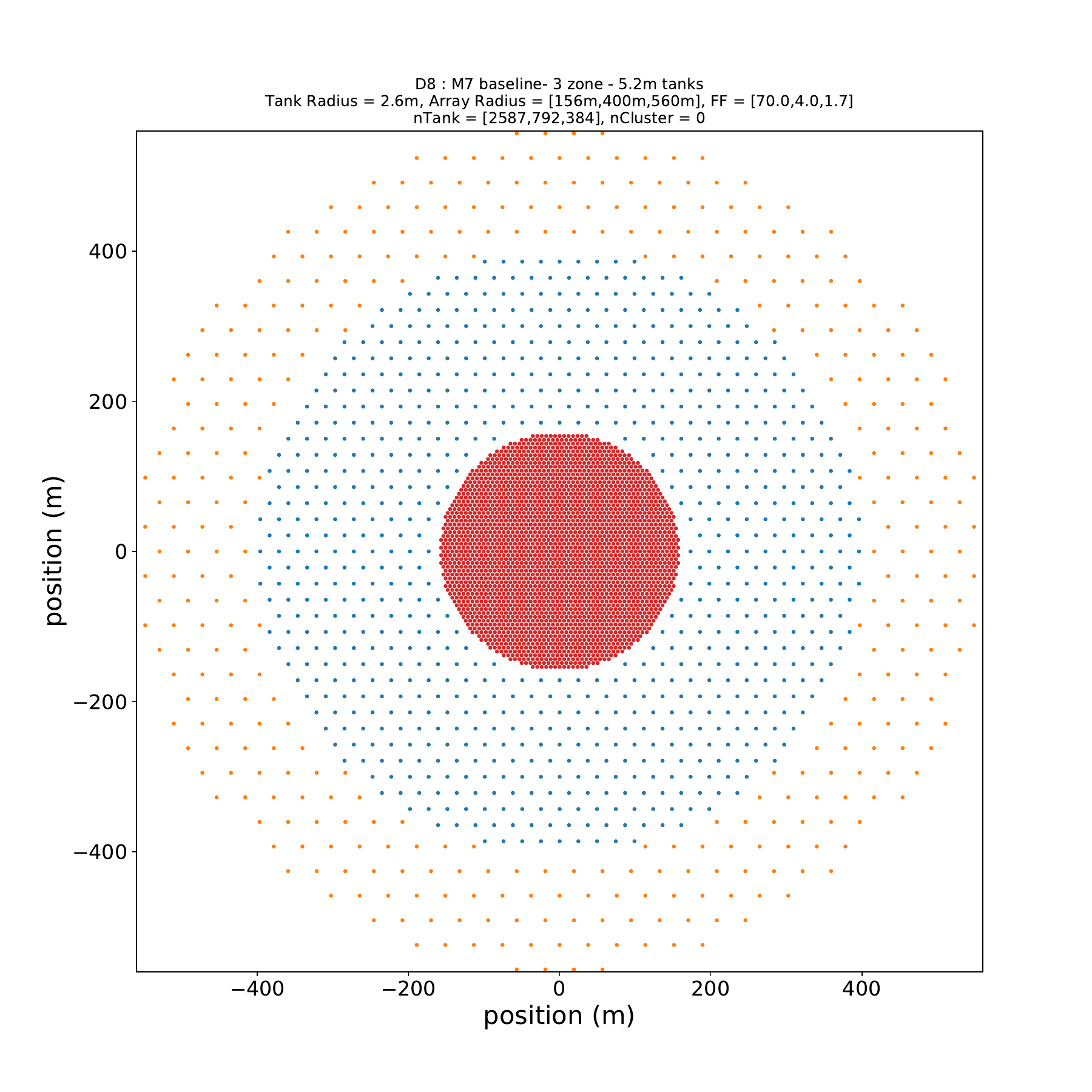}
  \caption{The baseline layout for SWGO consists of three zones with a decreasing fill factor. Each dot in the image represents a double-layer water-Cherenkov detector \cite{white_paper}.}
  \label{fig:layout}
\end{figure}

The Southern Wide-field Gamma-ray Observatory (SWGO) will feature double-layer water-Cherenkov detector units.
Each unit is a 5.2\,m wide and 4.2\,m high water tank with an optical separator at about a fifth of the tank height.
In the center of the tank, there are two photomultiplier tubes (PMTs), one facing upwards and one facing downwards into the optically separated lower layer.
The downwards-facing PMT allows for the identification of muons, which will easily penetrate the upper layer.
The baseline layout for SWGO, as shown in \autoref{fig:layout}, features a central zone in which the detector units are densely packed to achieve a good angular resolution.
The central zone is surrounded by two rings with decreasing fill factors, which extend the effective area and thus enhance high-energy sensitivity \cite{white_paper}.

The entire array will be subject to constant bombardment from secondary particles produced in cosmic ray induced extensive air showers.
Whenever charged particles traverse a detector unit, the unit will be read out by the local trigger system described in \autoref{sec:local}.
This results in a readout data rate of 14\,GiB per second.
Since most of the data is caused by background events, it is advisable to reduce the readout to a more reasonable amount.
To achieve this, an array trigger is required that only reads out the array if a gamma-ray induced air shower is hitting the array, see \autoref{sec:array}.

\section{Local trigger}
\label{sec:local}

\begin{figure}[htb]
  \centering
  \includegraphics[width=0.8\textwidth]{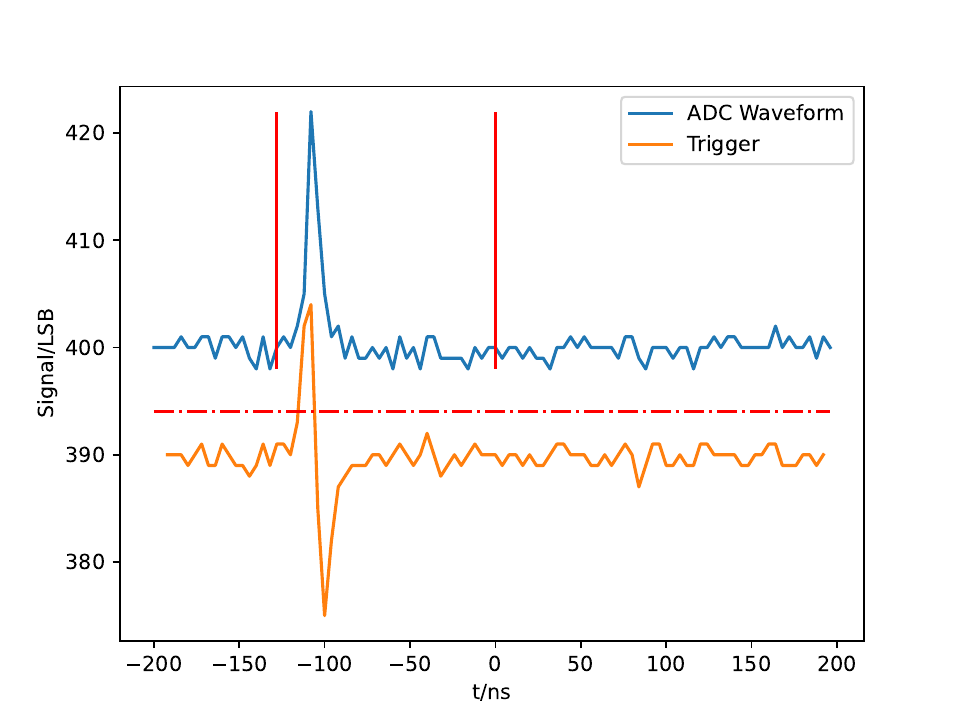}
  \caption{The ADC waveform (blue line, top) is convolved with the trigger kernel to produce the gradient (orange line, bottom). When the gradient exceeds the trigger threshold (red line, dash-dotted) a 128\,ns long trace around the trigger time is read out (vertical lines). The trigger lines are shifted upwards by 390\,LSB for visualization purpose.}
  \label{fig:local}
\end{figure}

Cherenkov light produced by relativistic charged particles in the water of the detector units creates photoelectrons (PEs) in the PMTs.
In the baseline design, the PMT signal is transferred out of the detector units and digitized remotely.
The readout electronics are realized in FPGA logic and sample with 12\,bit resolution and 250\,MHz sampling rate, resulting in 0.5\,GiB per second for each PMT.
To only read out the PMT signal when it detects PEs, a trigger kernel is applied to the digitized waveform of the PMT.
The kernel computes the gradient and allows it to trigger on the rising edge of the signal, as shown in \autoref{fig:local}.
Only the 128\,ns around the trigger time are read out and sent to the data acquisition servers \cite{tech_op}.

For the understanding of the data rate and interplay of local and array triggers, it is crucial to understand the readout electronics properly.
We developed a dedicated simulation framework for the PMT and the readout electronics that turns PEs into local triggers.
The simulation uses measured signal templates and considers PMT charge and time spectra as well as ADC phase jitter and noise.
It allows us to read in the detector simulations, which consist of lists of PE arrival times, and turn them into lists of local trigger times and their respective waveforms.
At the time of writing, this is the most realistic electronics simulation of SWGO available.

\section{Array trigger evaluation}
\label{sec:array}

An array trigger is an algorithm implemented in software that decides whether the array should be read out or not.
It defines time intervals in which all local triggers are stored to disk.
Those so-called array trigger intervals (ATIs) serve two purposes.
First, they reduce the amount of data that is written to disk.
Second, they define finite chunks of data that should contain a single air shower.
There are different strategies that can be pursued to achieve a good data reduction while not removing too much data to impact the performance of the observatory.
The following strategies will be further discussed in this work:
First, a simple multiplicity trigger that counts the number of local triggers within a moving time window.
If the number of local triggers exceeds a threshold, the array trigger activates.
As soon as the number drops below the threshold, the array trigger deactivates.
The active time with some padding is then used as ATI.
Second, to use the unique muon tagging capabilities of the detector units \cite{double_layer}, a multiplicity trigger strategy that only counts local triggers in the upper layer if there is no coinciding lower layer trigger in the same detector unit.
Third, a trigger strategy that weights the local triggers by the sum of their waveforms.
This is approximately proportional to the number of PEs and thus the deposited energy in the detector.
This trigger strategy uses the bottom layer to veto out the large signals produced by muons.

\begin{figure}[htb]
  \centering
  \includegraphics[width=0.8\textwidth]{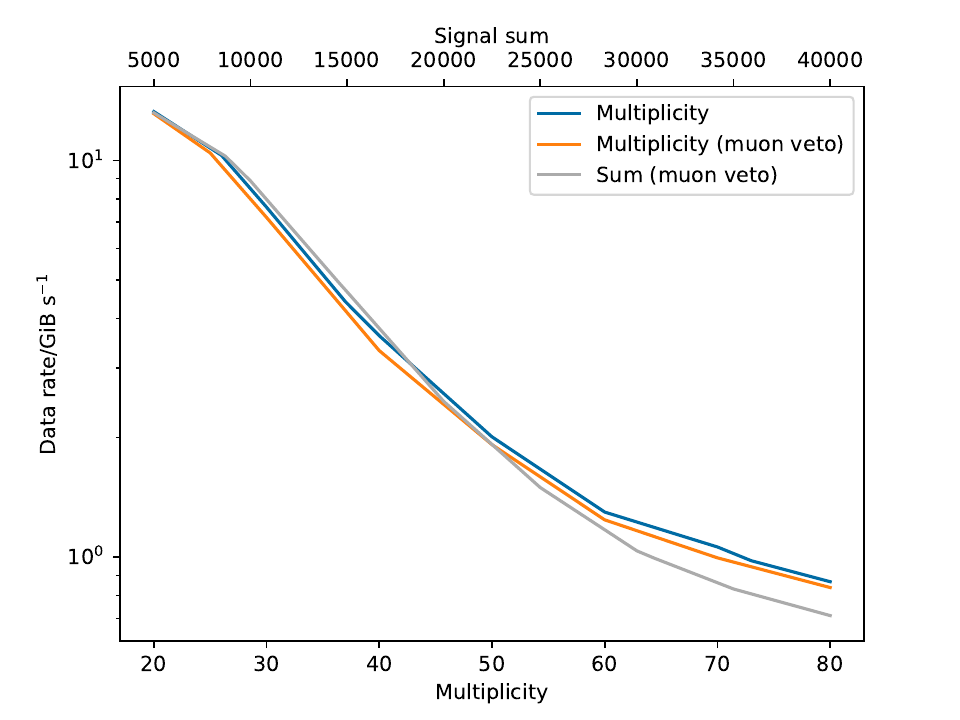}
  \caption{The readout data rate is plotted against the trigger threshold of the multiplicity trigger (bottom axis) and the threshold of the sum trigger (top axis).}
  \label{fig:rate}
\end{figure}

The relationship of the trigger thresholds to the data rate achieved by the respective trigger strategy is shown in \autoref{fig:rate}.
At low thresholds, all trigger strategies are active almost always, causing high data rates.
With increasing thresholds, the data rate is reduced.
We calculate the data rates by simulating the cosmic ray flux onto the observatory.
Simulations of extended air showers induced by primary protons with energies between 30\,GeV and 10\,PeV are passed through the detector simulation and our electronics simulation.
The cosmic ray showers are weighted by the cosmic ray flux and randomly selected over a certain time period.
This results in a continuous stream of local triggers that closely resemble the expected detector output of SWGO.
Afterward, the different array trigger strategies are applied to the stream of local triggers and their ATIs are evaluated.
Each local trigger within an ATI contributes to the data rate, while those outside of ATIs are ignored.
We repeat this process for multiple thresholds until we find a threshold that satisfies the data rate requirement of 2.5\,GiB/s.

To compare the performance of array trigger strategies, we add gamma-ray induced air showers to our simulation.
The gamma-ray showers are added on top of the cosmic ray showers to create a new stream of local triggers on which the array trigger strategies can be evaluated.
For each ATI, it is checked which air shower causes the most local triggers within the ATI.
If this shower is a gamma-ray shower that causes at least half of the local triggers, it is considered a triggering shower.
For each energy, it is possible to calculate the fraction of triggering showers to simulated gamma-ray showers.
This trigger fraction can be multiplied by the area over which the gamma-ray showers were simulated to get an effective area of the observatory.

\begin{figure}[htb]
  \centering
  \includegraphics[width=0.8\textwidth]{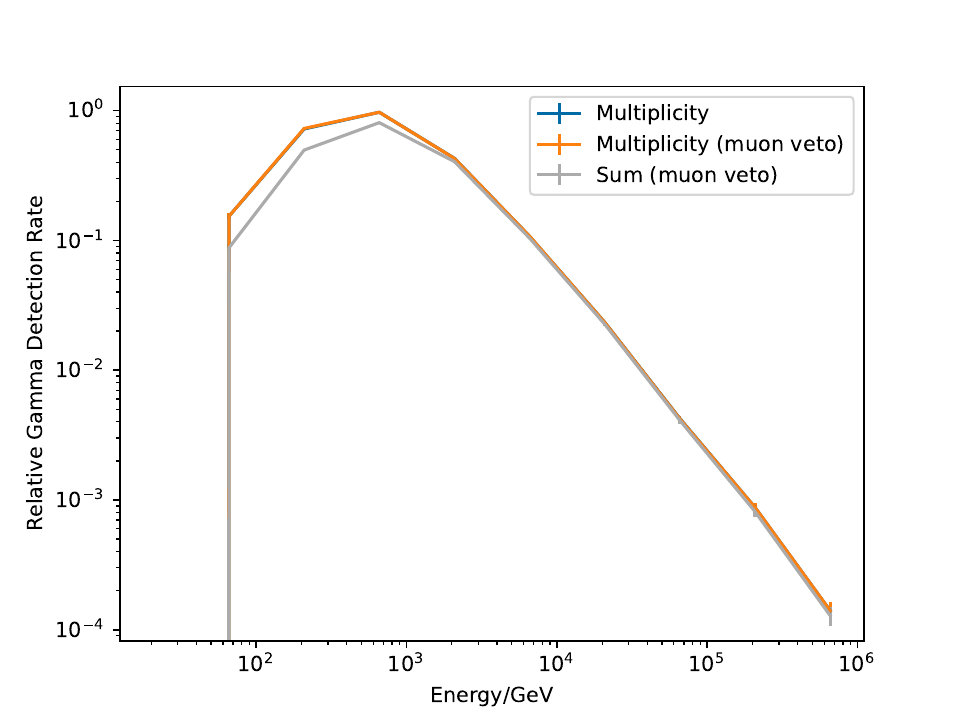}
  \caption{The detection rate per energy bin is plotted for the three array trigger strategies. For easier comparison, the curves are normalized so that the highest rate corresponds to one. The two multiplicity strategies (blue and orange lines) on the top overlap each other. The sum trigger strategy (gray line) has a slightly lower detection rate.}
  \label{fig:merit}
\end{figure}

For low energies, the fraction of gamma-ray showers that trigger and thus the effective area is getting smaller.
The effect on the detectability of a gamma-ray source can be estimated by multiplying the flux of the source by the effective area from the array trigger evaluation.
\autoref{fig:merit} shows the effective detection rate of gamma-rays from a typical source with a spectrum proportional to $E^{-2.5}$ for the different array trigger strategies.
For the high-energy bins, the detection rate closely follows the source spectrum.
For lower energies, the rate is reduced since most gamma showers are no longer able to trigger the array.
All trigger strategies show the same behavior with a peak detection rate at about half a TeV.

\begin{figure}[htb]
  \centering
  \includegraphics[width=0.8\textwidth]{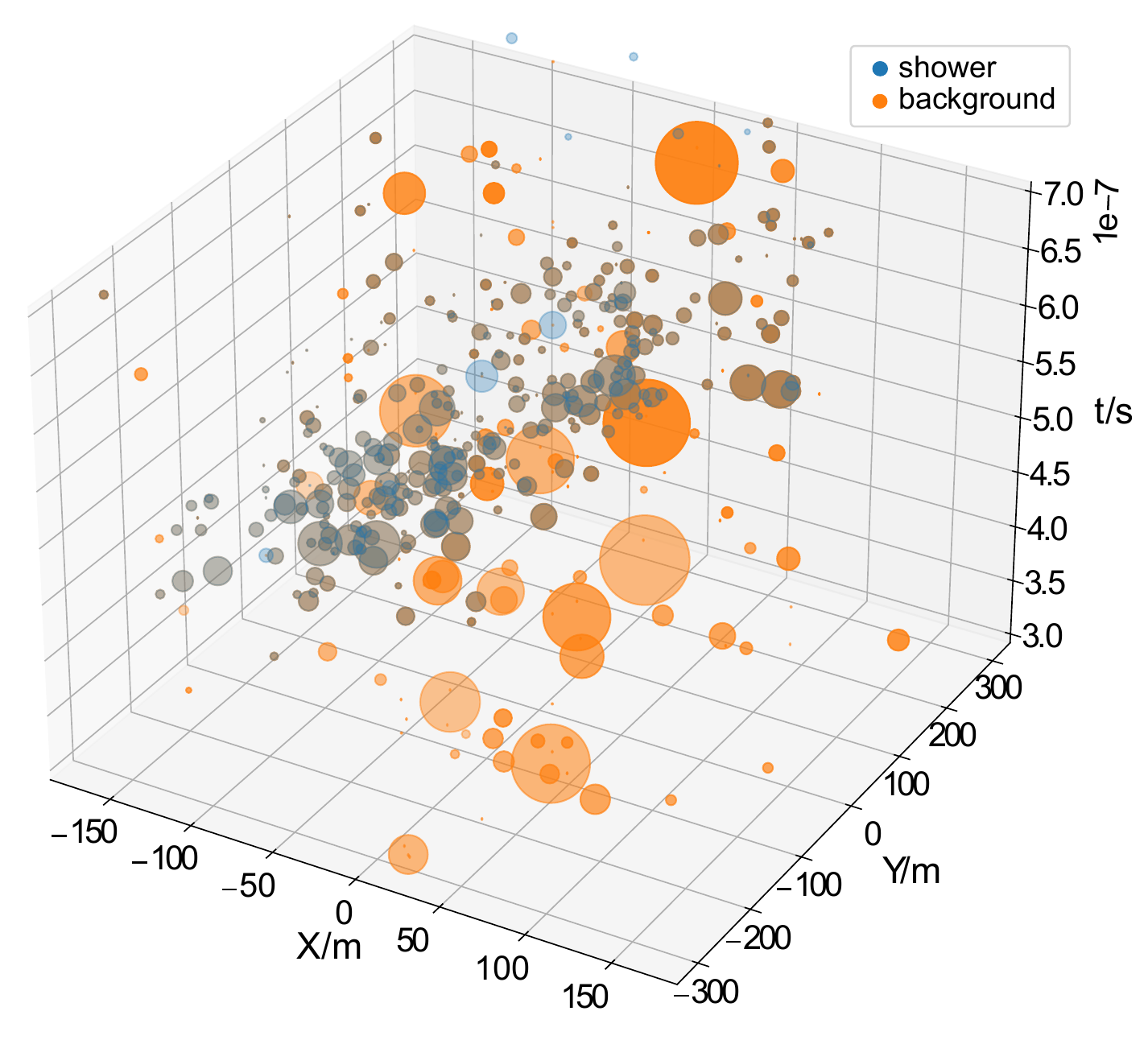}
  \caption{The gamma-ray shower (blue) is embedded in a cosmic ray background (orange). Each dot represents a local trigger at the position of its detector unit and its trigger time. The size of the dots is their integrated signal. It is evident, that the reconstruction of such a shower poses a serious challenge.}
  \label{fig:3d}
\end{figure}

\section{Reconstruction}
\label{sec:reco}

The discussion in the previous section did not consider the reconstruction of gamma-ray showers.
After the array trigger reads out an event, it has to be passed to the reconstruction.
At the time of writing, the official SWGO reconstruction chain does not support shower reconstruction in the presence of background events.
Therefore, it is not possible to evaluate the quality of the array trigger strategies in regard to the reconstruction performance.
Nonetheless, some qualitative statements can be made by looking at some ATIs, as seen in \autoref{fig:3d}.
Small air showers, as they are typical for low-energy gamma-rays, are embedded in an overwhelming cosmic ray background.
To identify and reconstruct such a shower, it is necessary to find the shower core and fit the shower plane.
A good array trigger strategy should assist in this endeavor by removing as much background as possible.
Additionally, showers that have no chance to be reconstructed should be dropped by the array trigger strategy to reduce the data rate.
This is a great source of inspiration for the design of future array trigger strategies.
They should consider directional information, as it is necessary for the plane fitting, and use it for their trigger decision.

\section{Outlook}

With the developed trigger simulation framework, we are able to test new array trigger strategies for SWGO.
As discussed in \autoref{sec:reco}, it is important to investigate the interplay between trigger and reconstruction further.
We are currently working on the support of a realistic background simulation for the official SWGO reconstruction chain to be able to evaluate array trigger strategies together with the reconstruction.
Furthermore, we plan to test different strategies to achieve the best array trigger performance.
We will focus on strategies that use knowledge about the structure of air showers to ensure that only high-quality events are read out.

\acknowledgments

The SWGO Collaboration acknowledges the support from the agencies and organizations listed here: \url{https://www.swgo.org/SWGOWiki/doku.php?id=acknowledgements}

\bibliographystyle{JHEP}
\bibliography{bib}

\end{document}